\documentstyle[12pt]{article}
\begin{document}
\title{Pygmy dipole resonances in relativistic\\
random phase approximation}
\author{D. Vretenar$^{1,2}$, N. Paar$^{1}$, P. Ring$^{1}$,
and G.A. Lalazissis$^{1,3}$
\vspace{0.5 cm}\\
$^{1}$ Physik-Department der Technischen Universit\"at M\"unchen,\\
D-85748 Garching, Germany\\
$^{2}$ Physics Department, Faculty of Science, University of Zagreb\\
10000 Zagreb, Croatia\\
$^{3}$ Physics Department, Aristotle University of Thessaloniki,\\
Thessaloniki GR-54006, Greece}
\maketitle
\begin{abstract}
The isovector dipole response in $^{208}$Pb is described
in the framework of a fully self-consistent relativistic random 
phase approximation. The NL3 parameter set for the effective mean-field 
Lagrangian with nonlinear meson self-interaction terms, used 
in the present calculations, reproduces ground state properties 
as well as the excitation energies of giant resonances in nuclei. 
In addition to the isovector dipole resonance in $^{208}$Pb, 
the present analysis predicts the occurrence of low-lying 
E1 peaks in the energy region between 7 and 11 MeV. In particular,
a collective state has been identified whose dynamics correspond 
to that of a dipole pygmy resonance: the vibration of the excess
neutrons against the inert core composed of equal number of 
protons and neutrons.
\end{abstract}
\newpage
The multipole response of nuclei with large neutron excess has
been the subject of many theoretical studies in recent years. 
Exotic nuclei with extreme neutron to proton ratio exhibit many
interesting and unique structure phenomena. In addition to exotic
ground state properties, the onset of low-energy collective isovector
modes has been predicted and experimental evidence for these modes
in light nuclei has been reported~\cite{Aum.99}. Catara {\it et al.}
have studied the low-lying components in strength distributions 
of weakly bound neutron-rich nuclei~\cite{Cat.96}, and the effect of 
large neutron excess on the dipole response in the region of the giant
dipole resonance in O and Ca isotopes~\cite{Cat.97}. They have shown that
the neutron excess increases the fragmentation of the isovector giant
dipole resonance (GDR) and that the radial separation of proton and neutron 
densities leads to non-vanishing isoscalar transition densities to 
the GDR states. The fragmentation of the 
isoscalar and isovector monopole strength 
in neutron rich Ca isotopes has been studied in Ref.~\cite{HSZ.97}, and 
in Ref.~\cite{Cha.94} the onset of pygmy dipole resonances in Ca isotopes 
has been analyzed.

The pygmy dipole resonance, which is also the subject of the present 
study, results from the excess neutrons oscillating out of phase 
with a core composed of equal number of protons and neutrons. 
A number of theoretical models have been applied in studies of the 
dynamics of pygmy dipole resonances. These include: the three-fluid 
hydrodynamical model (the protons, the neutrons of the same orbitals as
protons, and the excess neutrons)~\cite{MDB.71}, the two-fluid (the core 
fluid and the neutron excess fluid) Steinwedel-Jensen hydrodynamical 
model~\cite{SIS.90}, density functional theory~\cite{Cha.94}, and
the Hartree-Fock plus random phase approximation (RPA) with Skyrme
forces~\cite{Cat.97,ACS.96}.
More recently, large scale shell model calculations have been 
performed in studies of pygmy and dipole states in O 
isotopes~\cite{SS.99}, and dipole and spin-dipole strength distributions 
in $^{11}$Li~\cite{SSB.00}.
  
There is also experimental evidence for possible pygmy dipole states
in $^{208}$Pb. Studies of the low energy spectrum by elastic photon 
scattering~\cite{SAC.82}, photoneutron~\cite{BCA.82}, and electron
scattering~\cite{Kuh.81} have detected fragmented E1 strength in the
energy region between 9 and 11 MeV. The fine structure exhausts 
between 3 and 6\% of the E1 sum rule. The relationship between 
coherent neutron particle-hole (p-h) 
excitations and the onset of dipole pygmy 
resonances in $^{208}$Pb  has been investigated in the Hartree-Fock
plus RPA model~\cite{ACS.96}. A concentration of strength has been 
found around 9 MeV exhausting 2.4\% of the E1 sum rule. In particular,
two pronounced peaks have been calculated at 8.7 MeV and 9.5 MeV, which 
appear as likely candidates to be identified as pygmy resonances.
The exact location of the calculated pygmy states will, of course, 
depend on the effective nuclear interaction. Therefore, it would be
important to compare the predictions of various nuclear effective 
forces with experimental data. $^{208}$Pb is a particularly good
example, since all nuclear structure models have been tested in the 
description of ground and excited state properties of this doubly
magic spherical nucleus. The pygmy dipole resonance can be directly
related to the neutron excess, and therefore the splitting between
the GDR and the pygmy resonance represents a measure of the neutron
skin. Precise information on neutron skin in heavy nuclei is essential
for the quantification of the isovector channel of effective nuclear
forces.

In the present study the isovector dipole response in $^{208}$Pb 
is described in the framework of a fully self-consistent relativistic 
random phase approximation (RRPA). The same effective Lagrangian generates 
the Dirac-Hartree single-particle spectrum and the residual particle-hole 
interaction. The RRPA response functions with
nonlinear meson terms have been derived in Refs.~\cite{ma.97},
and applied in studies of isoscalar and isovector giant resonances.
However, the model configuration space did not include the 
negative energy Dirac states, and therefore model calculations 
did not reproduce the results 
obtained with the time-dependent relativistic mean-field
model~\cite{VBR.95,Vre.97}. The contribution of pairs formed from 
occupied positive-energy states and empty negative-energy states
(in the $no-sea$ approximation), is essential for current conservation and 
the decoupling of the spurious state~\cite{Daw.90}. In addition, 
configurations which involve negative-energy states give an important
contribution to the collectivity of excited states. In a recent 
study~\cite{Vre.00} we have employed a fully self-consistent RRPA, 
including configuration spaces with negative-energy Dirac states,
to calculate the isoscalar dipole resonance structure in $^{208}$Pb.
Two basic isoscalar dipole modes have been identified, and the discrepancy
between the calculated strength distribution and current experimental data
has been analyzed.
 
In Fig.~1 we display the isovector dipole strength 
distribution in $^{208}$Pb 
(left panel), and the corresponding transition densities to the two states 
at 7.29 MeV and 12.95 MeV (right panel). 
The calculations have been performed within the framework of
self-consistent Dirac-Hartree plus relativistic RPA. The effective
mean-field Lagrangian contains nonlinear meson self-interaction terms,
and the configuration space includes both particle-hole pairs and pairs 
formed from hole states and negative-energy states. The discrete 
spectrum of RRPA states has been folded with a Lorentzian distribution 
with a width of 0.5 MeV.

The strength distribution has been calculated with the
NL3~\cite{LKR.96} parameter set for the effective mean-field 
Lagrangian. This force has been
extensively used in the description of a variety of properties
of finite nuclei, not only those along the valley of $\beta$-stability,
but also of exotic nuclei close to the particle drip lines. 
Properties calculated with NL3 indicate that this is probably
the best effective interaction so far, both for nuclei at and away from the
line of $\beta $-stability. 
In particular, in Ref.~\cite{Vre.97} it has been shown
that the NL3 ($K_{\rm nm} = 271.8$ MeV) effective interaction 
provides the best description of experimental data on 
isoscalar giant monopole resonances.
The calculated energy of the main peak in Fig.~1
$E_p = 12.95$ MeV has to be compared with the experimental value of the excitation energy of the isovector giant dipole resonance: 
$13.3\pm 0.1$ MeV~\cite{Rit.93}. In the energy region between 5 and 11 MeV
two prominent peaks are calculated: at 7.29 MeV and 10.10 MeV. In the 
following we will show that the lower peak can be identified as the 
pygmy dipole resonance. 

The transition densities to the states 
at 7.29 MeV and at 12.95 MeV are displayed in the right panel of 
Fig.~1. The proton and neutron contributions are shown separately;
the dotted line denotes the isovector transition density and the 
solid line has been used for the isoscalar transition density. As it 
has been also shown in Ref.~\cite{Cat.97}, although the isoscalar B(E1) 
to all states must identically vanish, the corresponding isoscalar 
transition densities to different states need not to be identically
zero. The transition densities for the main peak at 12.95 MeV 
display a radial dependence characteristic for the isovector giant
dipole resonance: the proton and neutron densities oscillate with 
opposite phases; the total isovector transition density is much 
larger than the isoscalar component; at large radii they both
have a similar radial dependence. A very different behavior is 
observed for the transition densities to the state at 7.29 MeV:
the proton and neutron densities in the interior region are not
out of phase; there is almost no contribution from the protons
in the surface region; the isoscalar transition density dominates
over the isovector one in the interior; 
the large neutron component in the surface
region contributes to the formation of a node in the isoscalar 
transition density. In Ref.~\cite{Cat.97} it has been shown that this 
last effect is also characteristic for very neutron-rich systems.
The transition densities to the state at 10.10 MeV display a radial
behavior which is intermediate between those shown in Fig.~1, but
closer to the isovector giant dipole at 12.95 MeV. We have also analyzed 
the RRPA amplitudes of the three states: the neutron p-h excitations 
contribute 65\%, and the proton 35\% to the total intensity of the 
isovector giant dipole at 12.95 MeV, while the neutron contribution 
is 86\% for the state at 7.29 MeV. The proton p-h excitations contribute
only 14\% to the total RPA intensity of this state. For the state 
at 10.10 MeV, on the other hand, we find 68\% of proton excitations
and only 32\% is the contribution from neutron p-h configurations.
However, 31\% of the total intensity comes from a single proton 
p-h state: $g 7/2^{-1}~ h 9/2$. We notice that in the study of 
neutron halos and E1 resonances in $^{208}$Pb~\cite{ACS.96}, performed
in the HF+RPA model with the SGII interaction, it was found that for the 
pygmy states the neutron response is a factor 10 larger than the proton 
response, whereas at energies corresponding to the GDR this ratio 
is about 1.6 or roughly N/Z.

The phenomenon of low-lying isovector dipole strength was already studied
almost thirty years ago in the framework of the three-fluid hydrodynamical 
model~\cite{MDB.71}. By using a generalization of the Steinwedel-Jensen
model~\cite{SJ.50} to three fluids:
the protons, the neutrons in the same orbitals as protons, and 
the excess neutrons, two normal modes of dipole vibrations
were identified: (i) vibrations of the protons against the two 
types of neutrons, and (ii) the vibration of the excess neutrons 
against the proton-neutron core. In the case of neutron-rich nuclei,
the later mode corresponds to pygmy resonances. For $^{208}$Pb, 
in addition to the GDR state at 13.3 MeV, a low-lying pygmy state 
at 4.4 MeV excitation energy was found in the analysis of
Ref.~\cite{MDB.71}. The dipole strength of this state, however, was 
negligible (2 orders of magnitude) compared to the GDR state.

In Fig.~2 we plot
the transition densities to the two states at 7.29 MeV and 10.10 MeV. 
The contributions of the excess neutrons ($82 < N \leq 126$) (solid), 
and of the proton-neutron core 
($Z,N \leq 82$) (dashed) are displayed separately. By comparing with 
the transition densities shown in Fig.~1, we notice that there is 
practically no contribution from the core neutrons ($N \leq 82$). 
The reason is, of course, that the p-h configurations which involve
core neutrons have much higher excitation energies.
For the GDR state at 12.95 MeV the transition densities of the 
excess neutrons and the core have the same sign in the interior
($r < 3 fm $), and opposite phases in the surface region. The overall
radial dependence is, however, very similar. The two transition
densities have the same sign for the state at 7.29 MeV. The core
contribution, however, vanishes for large $r$ and only oscillations 
of the excess neutrons are observed on the surface of $^{208}$Pb.

The difference in the collective dynamics of the two modes is also
exemplified in the study of transition currents. In Figs.~3 and 4
we plot the velocity fields for the peaks at 7.29 MeV and 12.95 MeV,
respectively. The velocity distributions are derived from the corresponding
transition currents, following the procedure described in Ref.~\cite{Serr.83}.
In both figures the velocity field of the proton-neutron core
($Z,N \leq 82$) (left panel), is separated from the contribution 
of the excess neutrons ($82 < N \leq 126$) (right panel). To the largest
velocity in Figs. 3 and 4 a vector of unit length is assigned. All the other
velocity vectors are normalized accordingly. We notice that both the 
core and the excess neutrons contribute to the velocity field of the 
giant resonance state (Fig.~4), though the largest velocities correspond 
to the vibrations of the excess neutrons in the surface region. For the 
state at 7.29 MeV, on the other hand, the core velocities are much smaller 
than those of the excess neutrons on the surface. The velocity fields in
Fig.~3 corroborate the picture of dipole pygmy resonances as oscillations
of the excess neutrons against the inert core of protons and neutrons in 
the same shell model orbitals. 

In conclusion, we have used a fully self-consistent relativistic random 
phase approximation to analyze the isovector dipole resonance structure
in $^{208}$Pb. In particular, we have investigated the relationship 
between the neutron particle-hole excitations at low energy and the 
onset of dipole pygmy resonances. For the effective interaction we 
have used the set of nonlinear parameters NL3. This interaction has 
been used in many recent applications of the relativistic mean field
model, both in stable nuclei and in nuclei far from the valley of
$\beta$-stability, including drip-line systems. In the particular 
case studied in this Letter, model calculations reproduce the excitation 
energy of the isovector giant dipole resonance in $^{208}$Pb. In addition, 
low-lying E1 peaks are calculated in the energy region between 
7 and 11 MeV . While some of these represent the fine structure
of the giant resonance, a collective state has been identified whose 
dynamics correspond to that of a dipole pygmy resonance. By analyzing
transition densities and velocity distributions, we have related the
onset of pygmy resonance at 7.29 MeV to the vibration of the excess
neutrons against the inert core composed of protons and neutrons
in the same shell model orbitals. Experimental information on the 
fragmentation of E1 strength in $^{208}$Pb is, however, only available
in the energy window $8 - 12$ MeV. In order to test the predictions of
the present analysis, it would be important to obtain experimental data 
also in the energy region below 8 MeV.  

\bigskip
\bigskip
\noindent
{\bf Acknowledgments}

This work has been supported in part by the
Bundesministerium f\"ur Bildung und Forschung under
project 06 TM 979, by the Deutsche Forschungsgemeinschaft,
and by the Gesellschaft f\" ur Schwerionenforschung (GSI) Darmstadt.

\newpage
{\bf Figure Captions}

\begin{itemize} 
\item{\bf Fig.1} Isovector dipole strength distribution in $^{208}$Pb 
(left panel),
and transition densities for the two peaks at 7.29 MeV and 12.95 MeV (right
panel). Both isoscalar and isvoector transition densities are displayed,
as well as the separate proton and neutron contributions. All transition
densities are multiplied by $r^2$.

\item{\bf Fig. 2} Isovector dipole transition densities to the 
7.29 MeV and 12.95 MeV RRPA states in $^{208}$Pb. 
The contributions of the excess 
neutrons ($82 < N \leq 126$) (solid), and of the proton-neutron core 
($Z,N \leq 82$) (dashed) are displayed separately. The transition
densities are multiplied by $r^2$.

\item{\bf Fig. 3} Velocity distributions for the RRPA state at
7.29 MeV in $^{208}$Pb. The velocity field of the proton-neutron core
($Z,N \leq 82$) (left panel), is separated from the contribution 
of the excess neutrons ($82 < N \leq 126$) (right panel).

\item{\bf Fig. 4} Same as in Fig. 3, but for the RRPA state at
12.95 MeV in $^{208}$Pb.
\end{itemize}
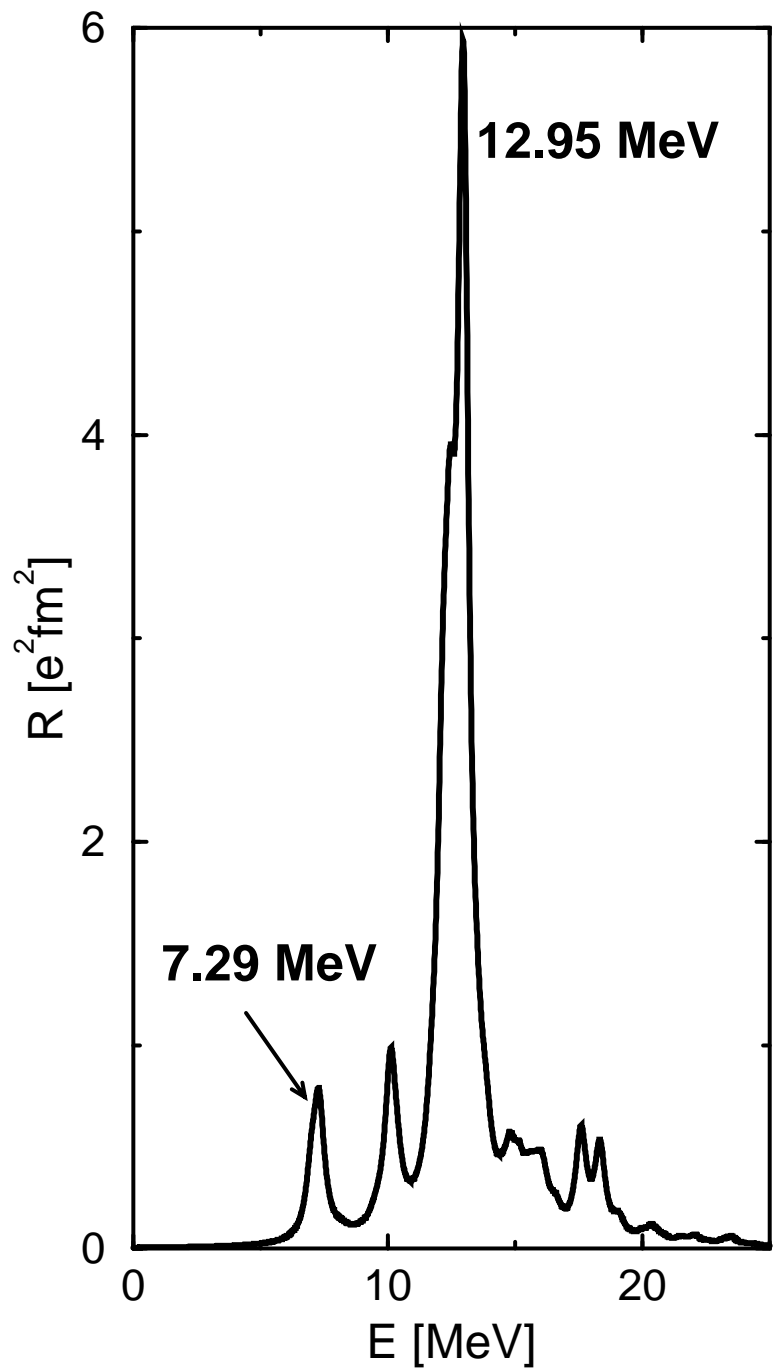
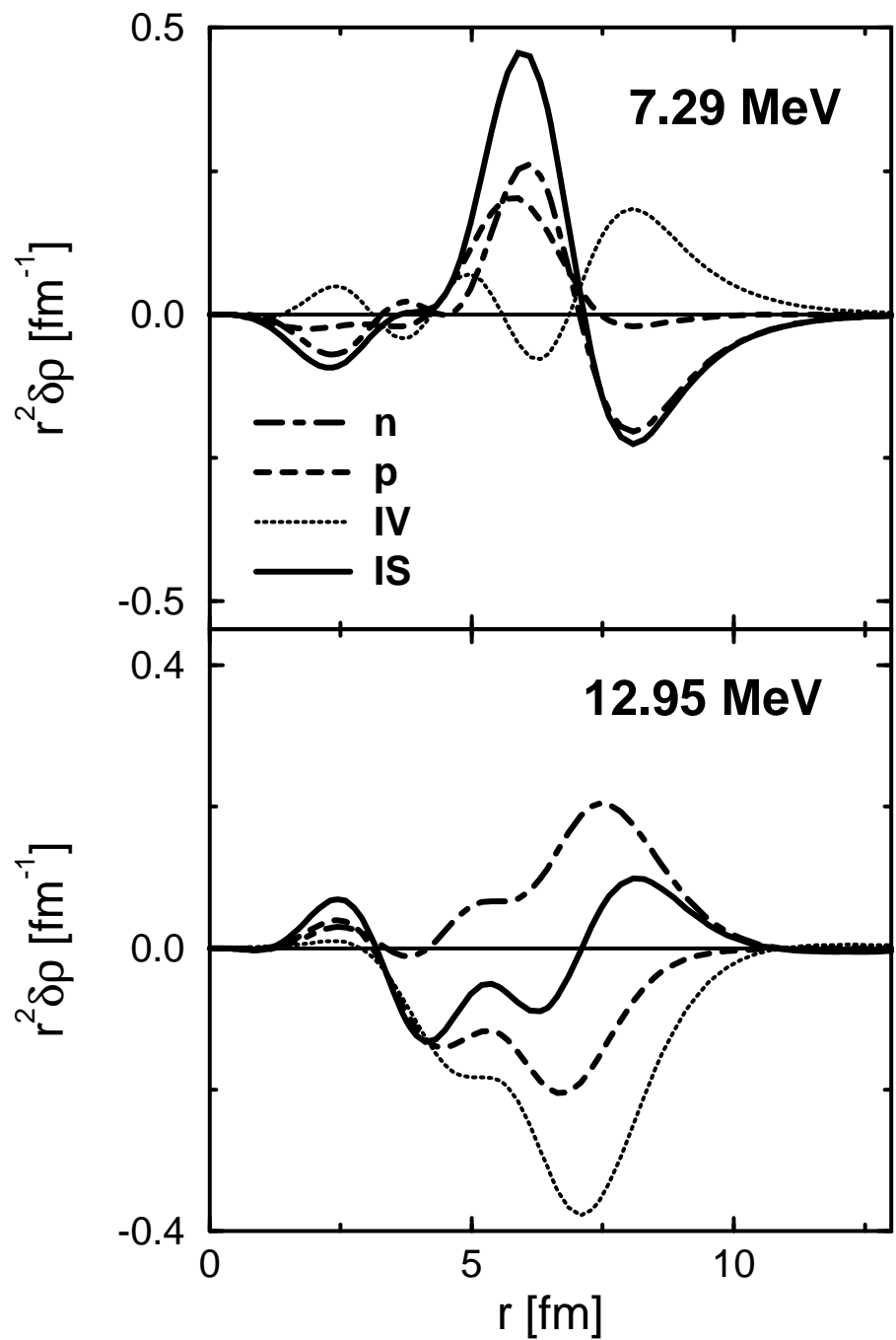
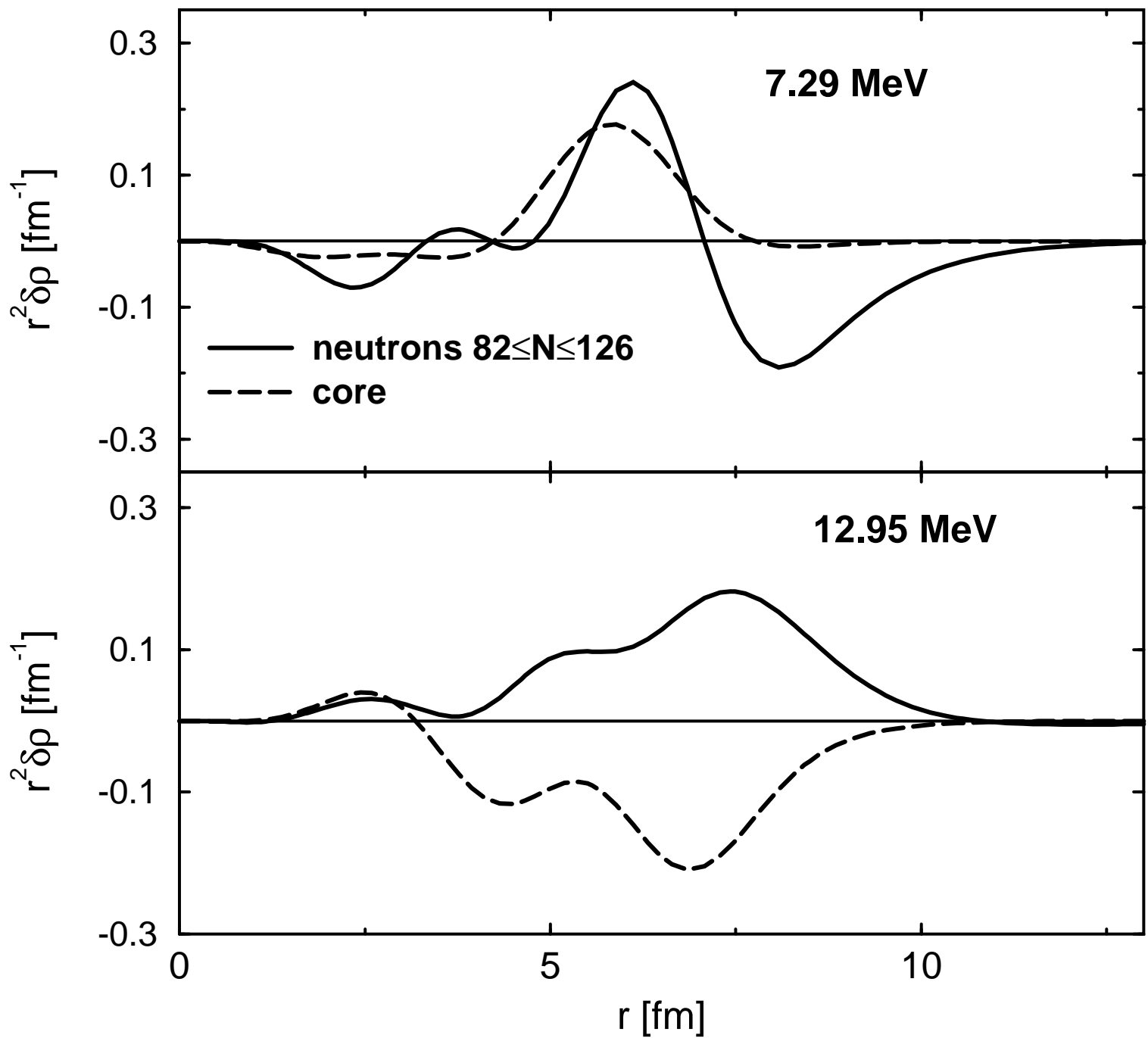
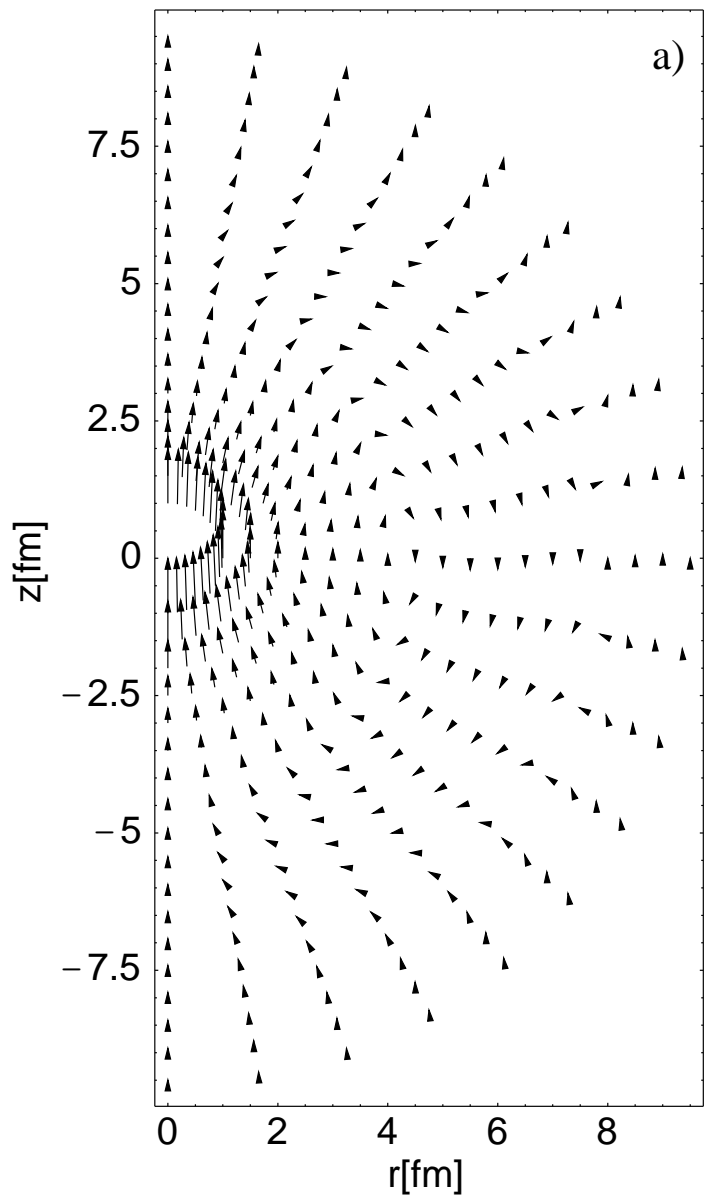
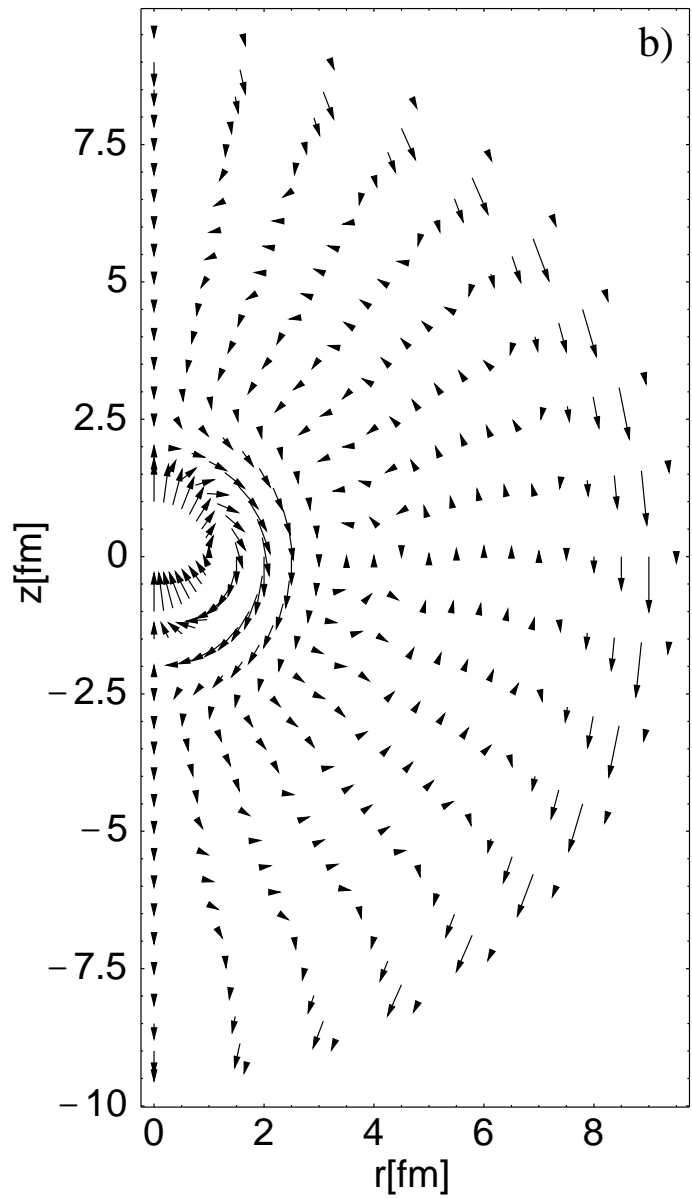
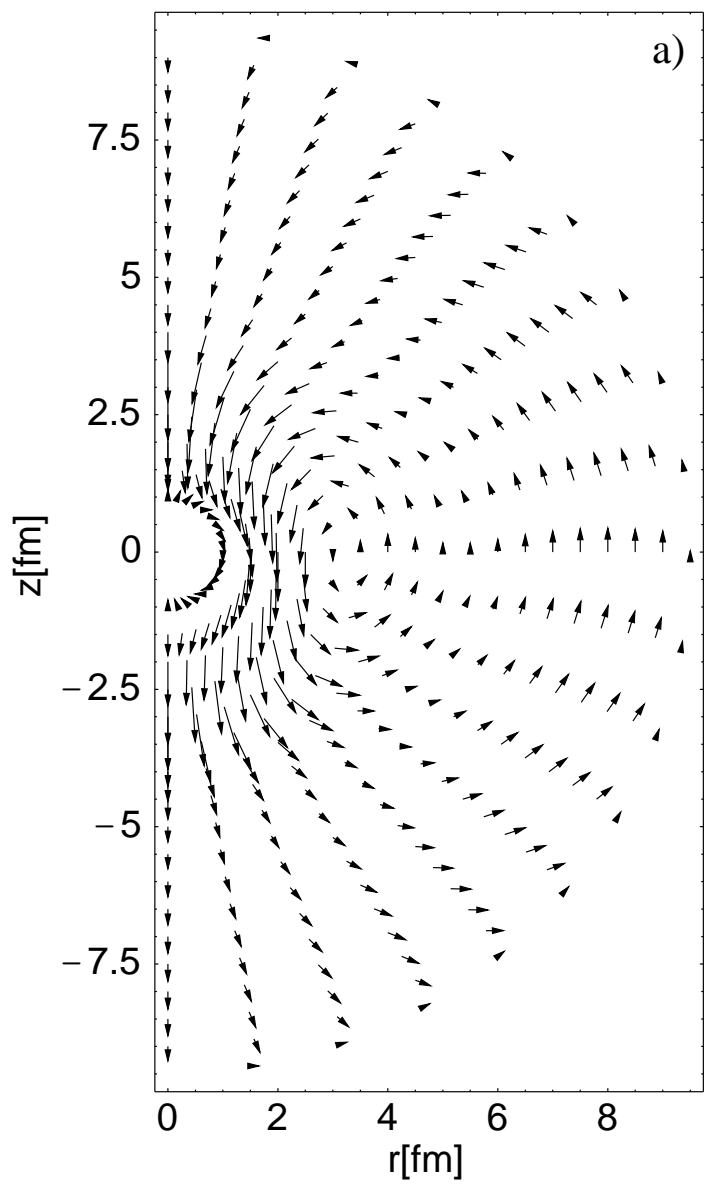
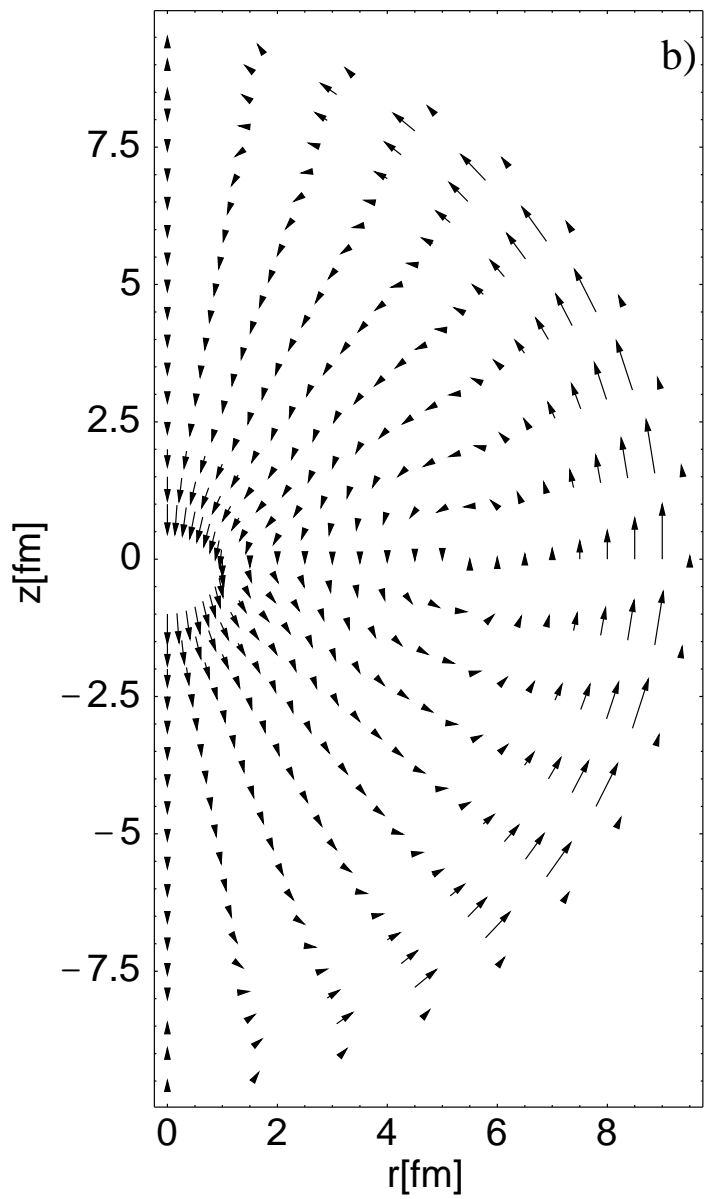
\end{document}